\documentclass[conference]{IEEEtran}
\IEEEoverridecommandlockouts

\usepackage{cite}
\usepackage{amsmath,amssymb,amsfonts}
\usepackage{algorithmic}
\usepackage{graphicx}
\usepackage{textcomp}
\usepackage{xcolor}
\usepackage{array}
\usepackage{multirow}

\def\BibTeX{{\rm B\kern-.05em{\sc i\kern-.025em b}\kern-.08em
    T\kern-.1667em\lower.7ex\hbox{E}\kern-.125emX}}
\begin{document}

\title{Toward Imagined Speech based Smart Communication System: Potential Applications on Metaverse Conditions\\

\thanks{This work was partly supported by Institute for Information \& Communications Technology Planning \& Evaluation (IITP) grant funded by the Korea government (MSIT) (No. 2017-0-00451, Development of BCI based Brain and Cognitive Computing Technology for Recognizing User’s Intentions using Deep Learning; No. 2015-0-00185, Development of Intelligent Pattern Recognition Softwares for Ambulatory Brain Computer Interface; No.2021-0-02068, Artificial Intelligence Innovation Hub).} }

\author{\IEEEauthorblockN{Seo-Hyun Lee}
\IEEEauthorblockA{\textit{Dept. Brain and Cognitive Engineering} \\
\textit{Korea University}\\
Seoul, Republic of Korea \\
seohyunlee@korea.ac.kr}
\and
\IEEEauthorblockN{Young-Eun Lee}
\IEEEauthorblockA{\textit{Dept. Brain and Cognitive Engineering} \\
\textit{Korea University}\\
Seoul, Republic of Korea \\
ye\_lee@korea.ac.kr}
\and
\IEEEauthorblockN{Seong-Whan Lee}
\IEEEauthorblockA{\textit{Dept. Artificial Intelligence} \\
\textit{Korea University}\\
Seoul, Republic of Korea \\
sw.lee@korea.ac.kr}
}

\maketitle

\begin{abstract}

Metaverse provides an alternative platform for human interaction in the virtual world. Since virtual platform holds few restrictions in changing the surrounding environments or the appearance of the avatars, it can serve as a platform that reflects human thoughts or even dreams at least in the metaverse world. When it is merged together with the current brain-computer interface (BCI) technology, which enables system control via brain signals, a new paradigm of human interaction through mind may be established in the metaverse conditions. Recent BCI systems are aiming to provide user-friendly and intuitive means of communication using brain signals. Imagined speech has become an alternative neuro-paradigm for communicative BCI since it relies directly on a person's speech production process, rather than using speech-unrelated neural activity as the means of communication. In this paper, we propose a brain-to-speech (BTS) system for real-world smart communication using brain signals. Also, we show a demonstration of imagined speech based smart home control through communication with a virtual assistant, which can be one of the future applications of brain-metaverse system. We performed pseudo-online analysis using imagined speech electroencephalography data of nine subjects to investigate the potential use of virtual BTS system in the real-world. Average accuracy of 46.54 ± 9.37 \% (chance level = 7.7 \%) and 75.56 ± 5.83 \% (chance level = 50 \%) was acquired in the thirteen-class and binary pseudo-online analysis, respectively. Our results support the potential of imagined speech based smart communication to be applied in the metaverse world. 
\end{abstract}


\begin{figure*}[t]
\centering
    \includegraphics[width=\textwidth]{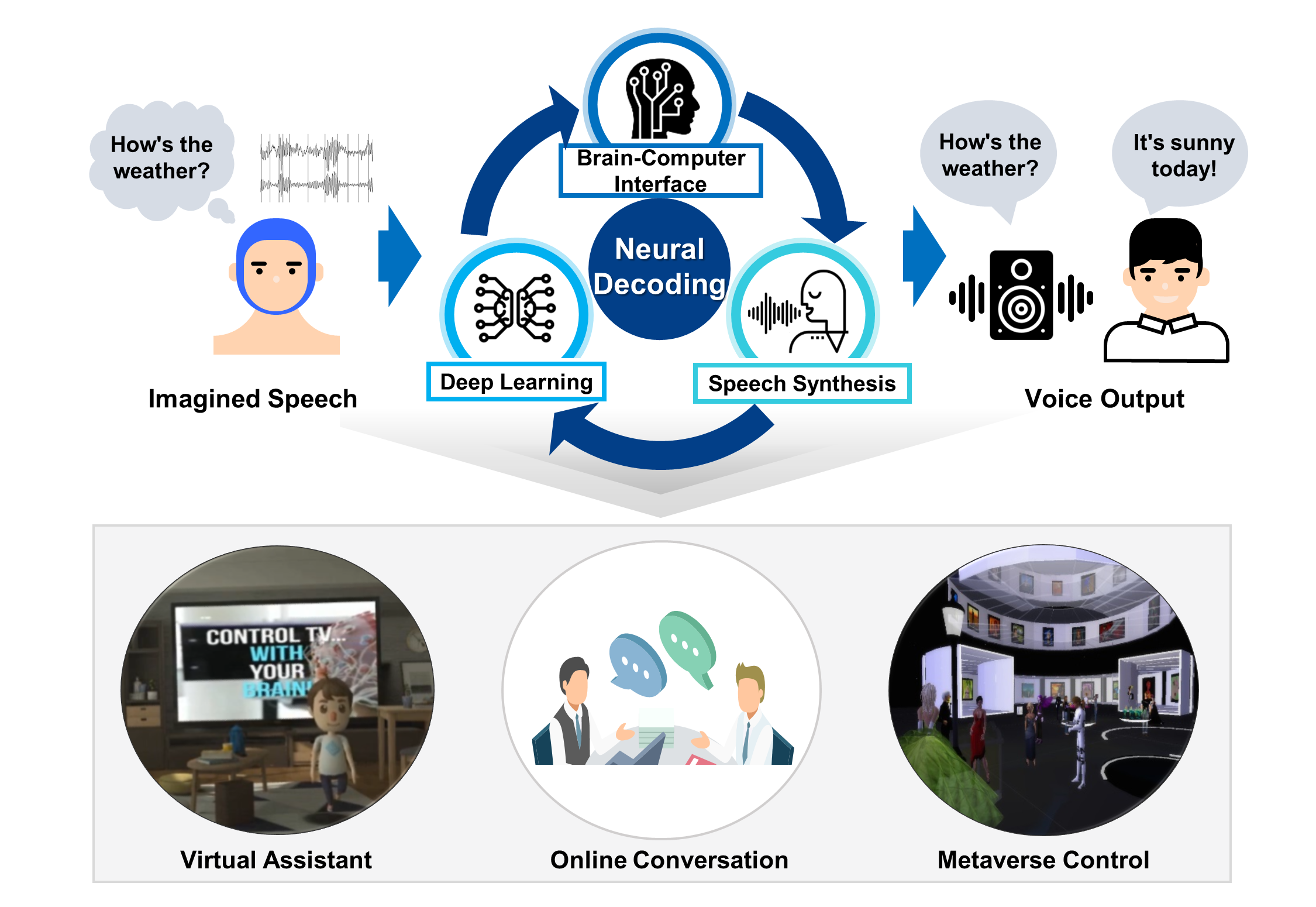}
    \caption{BTS communication system and potential applications to metaverse conditions.}
    \label{fig1}
\end{figure*}

\section{Introduction}

Metaverse is currently gaining attention as an alternative platform for human interaction in the virtual world \cite{dionisio20133d, luu2017real}. In parallel, brain-computer interface (BCI)\cite{wolpaw2000brain, wolpaw2002brain} is a technology to convey user intention by decoding brain signals of user's certain thoughts (e.g., imagined speech, imagined movements or visual imagery) \cite{lee2020neural, kevric2017comparison, lee2020spatio}. These two technologies may seem unrelated between each other, however, may create a new era of human interaction when elaborately combined. Since virtual platform holds few restrictions in changing the surrounding environments or the appearance of the avatars, it can serve as a platform to reflect human thoughts or even dreams in the metaverse world \cite{nevelsteen2018virtual}.

Imagined speech, which is a first person imagery of utterance without emitting an audible vocal output \cite{akbari2019towards}), is one of the emerging paradigms in the field of intuitive BCI\cite{nguyen2017inferring, lee2020neural} to deliver the commands or wishes to the virtual world. Conventional BCI paradigms such as steady-state visual evoked potentials and event-related potentials have shown robust performance in conveying user intention via brain signals \cite{lin2006frequency, lee2018high}, however, lack the intuitiveness to deliver user's direct thoughts because they require additional process (such as receiving stimulus or user training) \cite{lee2019towards}. Since imagined speech directly contains the word or sentence the user wants to say, it would be the most direct and convenient medium of BCI communication \cite{lee2020neural}. 
Similar to the current speech recognition systems that are mostly commercialized \cite{pratap2019wav2letter++}, imagined speech based communication system may be the future method to communicate or make commands only by imagining words \cite{lee2019towards, lee2019eeg}. Here, we call the imagined speech based communication system as brain-to-speech (BTS) system, which converts user's brain signal of imagined speech into an audible speech (see Fig. \ref{fig1}). When successfully decoded, this may hold an enormous potential to facilitate the BCI user's quality of life \cite{anumanchipalli2019speech}. 

BCI technology had originally developed to provide a communication method for people who has difficulties in communication with the outside world, such as, locked-in or paralyzed patients \cite{lee2015subject, lee2021functional}. However, it is currently expanding the range of its application to the healthy people as well, to provide convenience in their everyday life \cite{Lee2020decoding, lee2019eeg}. Together with the current development of virtual systems (such as virtual reality, augmented reality, and metaverse conditions) \cite{dionisio20133d}, BTS may be the next generation paradigm of neuro-communication. 

In this paper, we discuss the potential applications of BTS system as the future smart communication system, and present the potential applications of BTS system to the metaverse conditions. We performed a pseudo-online analysis of imagined speech to investigate the possibility of imagined speech based real-world communication system, and present a demonstration of a virtual assistant based smart home system as a potential applications of imagined speech based BTS system. Our results support the potential of imagined speech based smart communication to be applied in the metaverse world, and conversely, the virtual platform may act as an online training platform for practical BCI in the real world.

\begin{figure*}[t]
\centering
    \includegraphics[width=\textwidth]{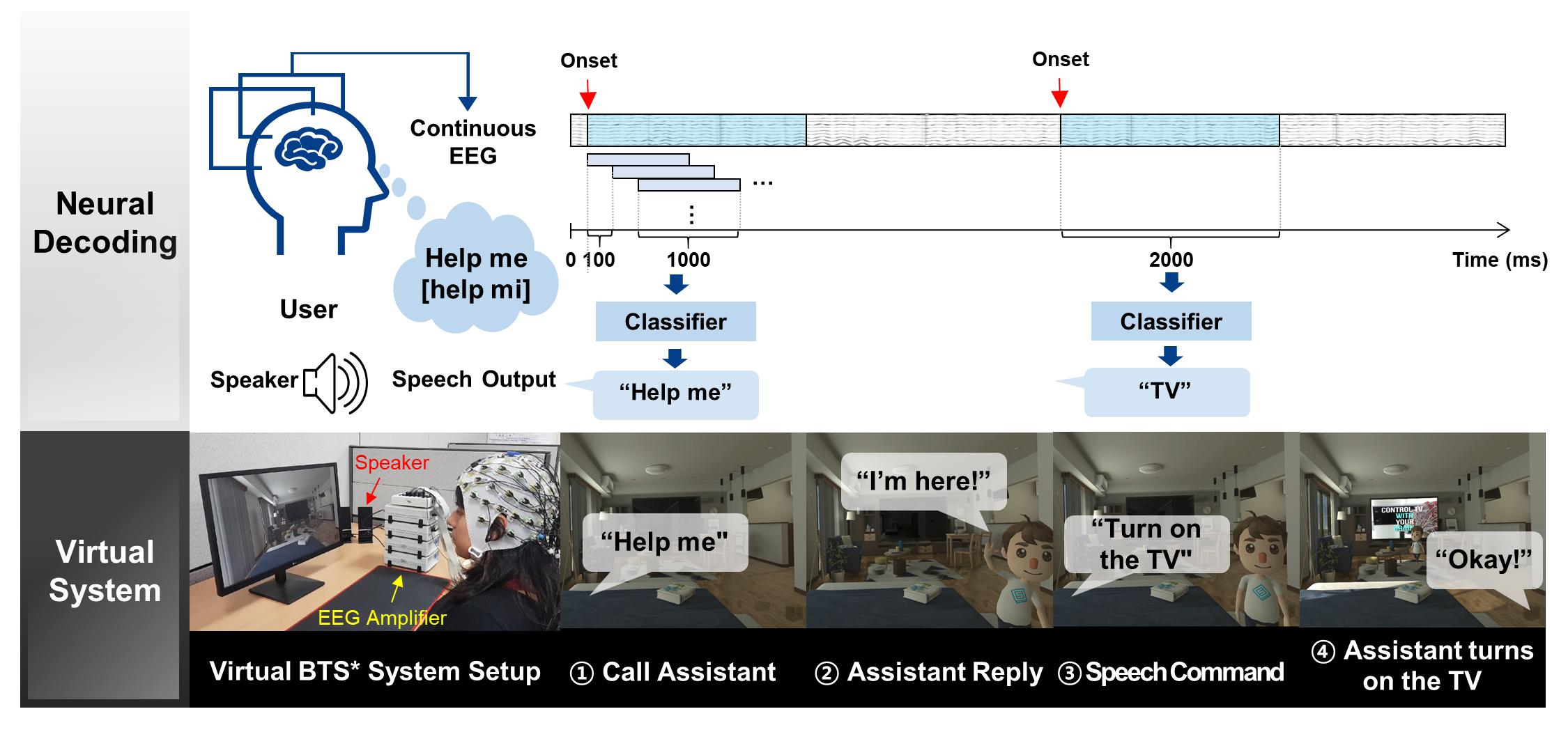}
    \caption{BTS system based on real-time decoding of imagined speech. Continuous EEG signal is processed in 1000 ms time window, with 100 ms shifting with 900 ms overlap. Using the calibrated classifier, each epoch was classified based on common spatial pattern and support vector classifier. The most frequent output of each 2000 ms epoch is selected and generated as an auditory output of the BTS system.}
    \label{fig2}
\end{figure*}

\section{Materials and Methods}
\subsection{Virtual BTS Platform}
Our virtual BTS platform was organized to suggest a potential application for the BTS system. We designed a scenario of making conversation with a virtual assistant in the smart home environment. If the user imagines to request help, the BCI system decodes the user intention to output the command "help me". Then the virtual assistant replies to offer help for the user in the virtual system. Although this may be a very basic and preliminary example of virtual BTS system, we believe enormous potentials exist to be expanded in the future. More detailed information about our imagined speech based virtual communication system is demonstrated in our demo video (http://deepbci.korea.ac.kr/intuitive-communication-system-using-imagined-speech/).

\subsection{Experimental Protocols}
EEG data of nine subjects (three males; age 25.00 ± 2.96) performing imagined speech of thirteen words/phrases (ambulance, clock, hello, help me, light, pain, stop, thank you, toilet, TV, water, and yes) were analyzed \cite{lee2020neural,lee2019eeg}. 100 trials per class for each subject were acquired using 64-channel EEG cap. Using these imagined speech EEG data, we performed pseudo-online analysis in order to investigate the potential use of BTS system in the real-world. The study was approved by the Korea University Institutional Review Board [KUIRB-2019-0143-01] and was conducted in accordance with the Declaration of Helsinki. Informed consent was obtained from all subjects. All experimental protocol are developed using BBCI toolbox\cite{krepki2007berlin}, OpenBMI\cite{leeMH2019eeg}, and EEGLAB\cite{lawhern2018eegnet}.

\subsection{Pseudo-online Analysis}
The epochs were divided randomly into two groups, 80 trials for the pre-training, and 20 trials for the pseudo-online test. Bandpass filtering of high-gamma frequency (30-120 Hz) was applied to each epoch \cite{lee2020neural}. Common spatial pattern (CSP) feature and support vector machine (SVM) classifier was trained using 80 trials of epochs in two conditions: 13-class classification and binary classification. For the binary classification we selected 'help me' and 'rest' class among the thirteen class words/phrases, since 'help me' was the first command to call the assistant in our proposed virtual BTS system. For the pseudo online analysis, continuous EEG signal was processed in 1000 ms time window, with 100 ms shifting with 900 ms overlap. Using the pre-trained classifier, each epoch was classified based on CSP feature and SVM classifier (see Fig. \ref{fig2}). In each 2000 ms epoch, the model generates single output as the command for the auditory output. In our model, the most frequent output contained in the 2000 ms epoch was selected and generated as an auditory output of the BTS system.

\begin{table}[!htbp]
    \centering
    \caption{Performance of decoding imagined speech of 13-class and 2-class for real-world communication}
    \begin{center}
    \begin{tabular}{>{\centering}p{0.2\columnwidth}|>{\centering}p{0.2\columnwidth}>{\centering\arraybackslash}p{0.2\columnwidth}}
    \hline
        ~ & \textbf{13-class} & \textbf{2-class}  \\ \hline
        Subject1 & 41.54 & 72.50  \\ 
        Subject2 & 45.00 & 85.00  \\ 
        Subject3 & 30.77 & 80.00  \\ 
        Subject4 & 63.46 & 77.50  \\ 
        Subject5 & 48.85 & 77.50  \\ 
        Subject6 & 56.92 & 77.50  \\ 
        Subject7 & 42.31 & 65.00  \\ 
        Subject8 & 43.46 & 70.00  \\ 
        Subject9 & 46.54 & 75.00  \\ \hline
        Avg. & 46.54 & 75.56  \\ 
        Std. & 9.37 & 5.83  \\ \hline
    \end{tabular}
    \end{center}
\end{table}

\section{Results and Discussion}
\subsection{Decoding Performance}
In the pseudo-online analysis we acquired average accuracy of 46.54 ± 9.37 \% (chance level = 7.7 \%) and 75.56 ± 5.83 \% (chance level = 50 \%) in the thirteen-class and binary pseudo-online analysis, respectively (see Table I). Our pseudo-online research using EEG of multiclass imagined speech implies further potentials for imagined speech based communication system. The proposed BTS system and virtual training platform may provide a useful tool for the development of intuitive BCI communication system. In real-time condition the classification result can be generated as a sound output in the BTS system, therefore, the user might feel as though his thought is spoken out \cite{lee2020neural}. 

\subsection{Potential of Imagined Speech Onset Detection}
\begin{figure}[t]
\centering
    \includegraphics[width=\columnwidth]{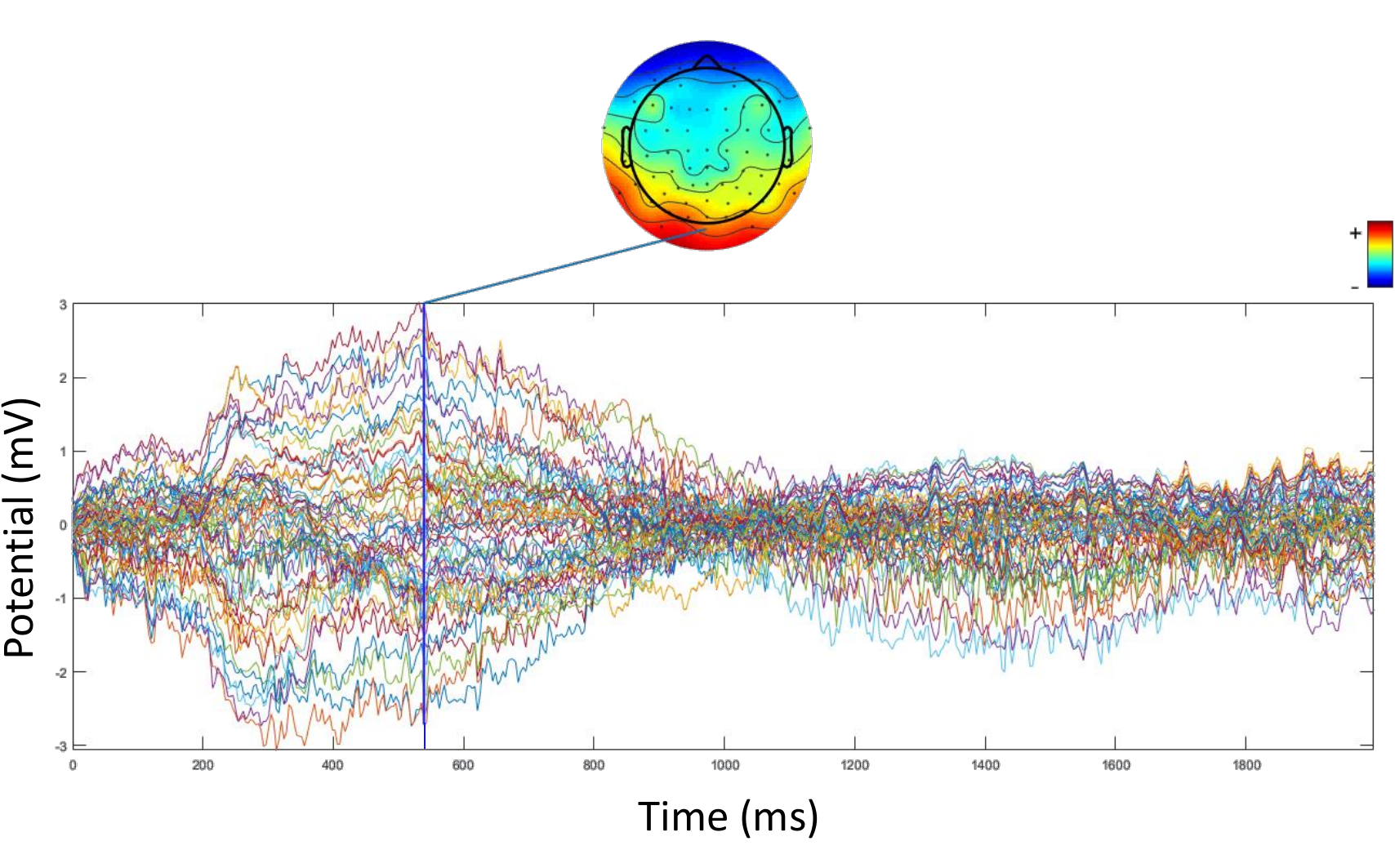}
    \caption{Example of potential changes for one epoch in single subject. The amplitude shows significant increase on the speech onset.}
    \label{fig3}
\end{figure}

In online BCI system, analyzing the onset of the imagery is a significant issue \cite{nguyen2017inferring}. In this research, we had a visual and rhythmic cue for the subject to perform imagery, however, we analyzed the potential changes in the EEG signal on the onset of the imagery. As shown in the Fig. \ref{fig3}, the amplitude of each channels have shown a significant variation on the speech onset point, compared to imagined speech performing intervals. Further analysis is required, however, this may imply the possibility of onset detection based on amplitude changes. Further analysis of onset detection may play a significant role in processing real-time BTS system.

\subsection{Future Applications for Metaverse Conditions}
Although our analysis was currently a pseudo-online analysis for virtual platform, this may be expanded widely to be adjusted for the metaverse conditions. BTS system may be used as a chatting platform in the online conversation, or controlling avatar in the metaverse conditions. Like our virtual environment, this can be applied to control virtual assistant or real-world smart home system with imagined speech commands. In addition, the suggested virtual platform may be used as an online training platform for neurofeedback to improve the user's BCI performance. Therefore, the virtual BTS system in this paper suggests a possible form of real-world application and online training platform.

\section{Conclusion}
In the BTS system, the decoded output is generated into a real speech sound output, therefore, the user may feel as though the imagined words are spoken out by someone else. Besides the classification issue, development in the current speech synthesis technology is enabling to directly synthesize audible speech from brain signals \cite{anumanchipalli2019speech}. The virtual BTS system in this paper shows the possible form of real-world application and online training platform. Our pseudo-online research using EEG of multiclass imagined speech displays potential of online BCI communication system operated using multi commands. The proposed BTS system and virtual training platform may provide a useful tool for the development of intuitive BCI communication system.

\bibliographystyle{IEEEtran}
\bibliography{mybib}

\vspace{12pt}
\vspace{12pt}

\end{document}